\newcommand{\diracslash}[1]{#1\llap{/\kern2pt}}

\newcommand{\be}{\begin{equation}}
\newcommand{\ee}{\end{equation}}
\newcommand{\bea}{\begin{eqnarray}}
\newcommand{\eea}{\end{eqnarray}}
\newcommand{\ba}[1]{\begin{array}{#1}}
\newcommand{\ea}{\end{array}}

\newcommand{\bt}{\begin{tabular}}
\newcommand{\et}{\end{tabular}}

\newcommand{\beas}{\begin{eqnarray*}}
\newcommand{\eeas}{\end{eqnarray*}}

\RequirePackage{lineno}
\documentclass[preprint,prd,aps,floats,nofootinbib,floatfix]{revtex4}

\DeclareSymbolFont{rsfs}{U}{rsfs}{m}{n}
\DeclareSymbolFontAlphabet{\mathrsfs}{rsfs}

\usepackage{graphicx}
\usepackage[mathscr]{eucal}
\usepackage{multirow}
\usepackage{graphicx,epstopdf}

\usepackage{graphicx}
\usepackage{subcaption}
\usepackage[utf8]{inputenc}
\usepackage[T1]{fontenc}
\usepackage{csquotes}
\usepackage[english]{babel}
\usepackage{hyperref}
\usepackage{upgreek}
\hypersetup{
    colorlinks=true,
    linkcolor=blue,
    filecolor=magenta,     
    citecolor=blue, 
    urlcolor=blue,
}
 
\usepackage{cleveref}

\begin{document}
\title{Impact of non-zero strangeness on thermodynamics of finite volume quark matter} 
 
 

\author{Nisha Chahal}\email{nishachahal137@gmail.com}
\author{Suneel Dutt}\email{dutts@nitj.ac.in}
\author{Arvind Kumar}\email{ kumara@nitj.ac.in}
\address{Department of Physics, Dr. B R Ambedkar National Institute of Technology, Jalandhar - 144008, Punjab, India}

\def\be{\begin{equation}}
\def\ee{\end{equation}}
\def\bearr{\begin{eqnarray}}
\def\eearr{\end{eqnarray}}
\def\zbf#1{{\bf {#1}}}
\def\bfm#1{\mbox{\boldmath $#1$}}
\def\hf{\frac{1}{2}}
\def\kp{\zbf k+\frac{\zbf q}{2}}
\def\km{-\zbf k+\frac{\zbf q}{2}}
\def\hwo{\hat\omega_1}
\def\hwt{\hat\omega_2}

\begin{abstract}
This paper investigates the impact of strangeness chemical potential and finite volume on QCD critical end point by employing a (2+1) flavored Polyakov quark meson model. Within the mean-field approximation, the model has been extended to study the effect of vector interactions on the thermodynamics of isospin-asymmetric quark matter. Susceptibilities of conserved charges like the quark and strangeness number are analyzed using Taylor's series expansion. The chiral phase transition boundary in the QCD phase diagram is found to be shifted towards higher values of the quark chemical potential ($\mu_q$) and lower temperature (T) for decreasing system size. On the other hand, there is an opposite change to lower quark chemical potential and higher temperature for decreasing strangeness chemical potential. 
\end{abstract}

\maketitle

\maketitle

\section{Introduction}
\label{intro}

Thermodynamics of strongly interacting matter, such as quark-gluon plasma (QGP), is of tremendous interest to theoretical and experimental physicists \cite{pandav2022search,ohnishi2012phase}. Heavy ion collision experiments like the Large Hadron Collider (LHC) \cite{bruning2012large} at CERN in Switzerland and the Relativistic Heavy Ion Collider (RHIC) at Brookhaven National Laboratory \cite{odyniec2010rhic,kumar2013star} are crucial in recreating the conditions of the early universe to understand the properties of QGP better. In addition to the experimental facilities, lattice QCD simulations is also a powerful technique for calculating the QCD partition function on a discrete space-time lattice \cite{ratti2018lattice}. It is a non-perturbative application of field theory based on the Feynman path integral technique and can also be used to study some important hadron properties \cite{gupta1999lattice, detar2009qcd}. The pressure, energy density, and susceptibilities of conserved charges can be obtained using lattice QCD simulations of thermodynamic phenomena. At high temperatures and low baryonic densities, these computations have anticipated a crossover transition from the confined to the deconfined state \cite{guenther2021overview}. On the other hand, due to the sign problem at higher baryonic chemical potentials and low-temperature values, the phase fluctuations derived from the complex fermion determinant are found to be very large \cite{muroya2003lattice}.

After formation, the QGP expands rapidly and cools down to form hadrons in a small region of space. Due to the finite size of colliding nuclei and the geometry of the collision, the fireball created in the initial stage is a finite-sized system. In this case, the finite size refers to the influence of the QGP's finite volume on its properties and evolution. Finite-size effects can affect particle flow, hadronization, and other variables sensitive to QGP properties. As a result, the phase transition line and the QCD critical point are modified due to finite volume considerations \cite{Braun:2011iz, MataCarrizal:2022gdr, chahal2023effects}. For low chemical potential values, the impact of finite size effects on the curvature of the chiral transition boundary has also been studied in lattice QCD \cite{de2007chiral,karsch2004chiral}.

The effective models serve as an essential tool to study the properties of matter formed in heavy-ion collisions at finite baryonic chemical potential values \cite{tripolt2014effect,palhares2010finite,magdy2017influence,magdy2019influence}. Using the quark-meson model, which serves as a low-energy model for QCD, the effects of finite volume and long-range fluctuations have been evaluated \cite{Braun:2010vd,article}. Based on the pion mass, it had been observed that the curvature continually decreases for periodic and anti-periodic boundary conditions as a function of system size, thus indicating that the phase transition boundary may shift in a finite volume. The earlier study of the Polyakov quark meson (PQM) model showed that the phase transition boundary remains just a crossover for very small systems \cite{bhattacharyya2017polyakov}. The effect of finite volume on the QCD phase diagram has also been analyzed considering different boundary conditions, such as stationary, periodic, and antiperiodic for cubic and spherical regions \cite{mata2022effects}. 
Also, the concept of finite strangeness chemical potential in quark matter is critical to our understanding of the fundamental building blocks of the universe. Incorporating finite strangeness chemical potential into the study of quark matter allows us to investigate the behavior of strange quarks in this exotic condition. The large value of strangeness has been shown to affect the generation of the neutron twin-stars, which can predict the existence of first-order phase transition at lower temperature and higher density values \cite{dexheimer2015role}. In the framework of the Hadron resonance gas (HRG) model, it has been highlighted that the critical temperature decreases with the increase in the value of strangeness chemical potential \cite{toublan2005qcd}. Hence, the quark-hadron phase boundary is modified as a consequence of finite strangeness chemical potential considerations and is of great importance in studying thermodynamic properties of quark matter \cite{mukherjee2019effects}.

In the investigation of QCD matter, the susceptibilities of conserved charges are particularly significant observables \cite{fan2019probing,chahal2022quark,chatterjee2012fluctuations,borsanyi2012fluctuations}. These have been recognized as observables that can be theoretically and experimentally estimated to better understand the critical end point (CEP). Studying the susceptibilities of conserved charges in finite volumes has gained popularity recently since it can shed light on the nature of QCD phase transitions and the equation of state of QCD matter in practical experimental settings \cite{bhattacharyya2014fluctuations}. A sizable shift in the chiral phase boundary has been reported in finite volume studies of earlier PQM model \cite{magdy2019influence}. Including isospin chemical potential and vector interactions in the PQM model have been found to affect the positioning of CEP significantly \cite{chahal2022quark}. In the present work, we use the three-flavored Polyakov quark-meson model, extended by the inclusion of vector interactions and isospin asymmetry, to study the thermodynamics of quark matter for finite system size and non-zero value of strangeness chemical potential.

This paper is organized as follows: In Sect.~\ref{method}, we have described the PQM model in detail and derived the grand canonical potential. In Sect.~\ref{results}, the impact of the non-zero value of strangeness chemical potential and finite volume on various thermodynamic variables and the QCD phase diagram has been discussed. In Sect.~\ref{summary}, the critical findings of the current work have been summarized.

\section{Polyakov quark meson model} \label{method}

The chiral quark meson model is an effective approach to study the strong interactions between mesons and quarks. The spontaneous breaking of the chiral symmetry in vacuum is described by including a scalar field that represents the chiral condensate. It is a non-perturbative model based on the mean-field approximation to study the thermodynamic properties of the system.
The model has been used to study various phenomena in QCD, such as the properties of hadrons, the phase structure of QCD at finite temperature and density, and the properties of quark-gluon plasma created in heavy-ion collisions \cite{mintz2013phase,skokov2010meson}. It has also been used to study the behavior of matter under extreme conditions, such as the properties of neutron stars and the early universe \cite{zacchi2015compact,otto2020hybrid}. The total Lagrangian of the model for $N_f$ flavors is given by \cite{PhysRevD.90.085001,newvector}
\begin{eqnarray}\label{effective}
 	\mathcal{L}=  \bar{\Psi} i \gamma^{\mu}{\partial}_{\mu} \Psi 
+ \text{Tr}\left(\partial_{\mu} \upvarphi^{\dagger} \partial^{\mu} \upvarphi \right) - m^{2} \text{Tr}\left(\upvarphi^{\dagger} \upvarphi \right) -
  \lambda_{1}\left[\text{Tr}\left(\upvarphi^{\dagger} \upvarphi\right)\right]^{2} -
   \\ \nonumber
  \lambda_{2}\left[{Tr}\left(\upvarphi^{\dagger} \upvarphi\right)^{2}\right]+
 	 c\left(\text{det}(\upvarphi)+\text{det}\left(\upvarphi^{\dagger}\right)\right) + \text{Tr}\left[H\left(\upvarphi+\upvarphi^{\dagger}\right)\right] 
   \\  \nonumber
  + \mathcal{L}_{qm} -\frac{1}{4}\text{Tr}\left(V_{ \mu \nu }V^{ \mu \nu }\right)
 	 + \frac{m_1^2}{2}V_{a \mu} V_{\mu}^a .
 	\end{eqnarray}

In the above equation, $\Psi$ = $(u,d,s)$ is the quark spinor for $N_c$ = 3, color degrees of freedom and $\upvarphi$ = $T_a(\sigma_a + i \gamma_5 \pi_a)$, where $T_a = \lambda_a$/2 are Gell-Mann matrices. The first term in Eq.~\ref{effective} accounts for the kinetic energy of massless quarks. The next two terms describe scalar mesons kinetic energy and mass term contributions. The terms involving $\lambda_1$ and $\lambda_2$ are quartic interaction terms, followed by the determinant term, which corresponds to U(1)$_A$ anomaly in QCD vacuum \cite{t1976symmetry} and explicit symmetry-breaking terms defined by $H=T_{a}h_a$. Through these symmetry-breaking terms, the $\sigma$ meson has a finite vacuum expectation value (VEV) and, consequently a finite quark mass \cite{beisitzer2014equation}. The last two terms in the Lagrangian are incorporated to define the vector meson interaction with mesons. Due to $SU(3)_L \times SU(3)_R$ symmetry in the effective Lagrangian, the term representing the quark-meson interaction can be written as \cite{kovacs2016existence}
\begin{equation}\label{left}
 	\mathcal{L}_{q m}=g_{s}\left(\bar{\Psi}_{L} \upvarphi \Psi_{R}+\bar{\Psi}_{R} \upvarphi^{\dagger} \Psi_{L}\right)-g_{v}\left(\bar{\Psi}_{L} \gamma^{\mu} L_{\mu} \Psi_{L}+\bar{\Psi}_{R} \gamma^{\mu} R_{\mu} \Psi_{R}\right).
 	\end{equation}
Here, $g_v$ and $g_s$ represent the coupling constants for the vector and scalar mesons, respectively. In the above equation, $L_{\mu}$ and $R_{\mu}$ are defined in terms of pseudovector $(A_{\mu}^a)$ and vector $(V_{\mu}^a)$ mesons as
\begin{eqnarray}
 L_{\mu}\left(R_{\mu}\right)=\frac{1}{2 \sqrt{2}}\left(\begin{array}{ccc}
\frac{\omega+\rho^{0}}{\sqrt{2}} & \rho^{+} & K^{\star+} \\
\rho^{-} & \frac{\omega-\rho^{0}}{\sqrt{2}} & K^{\star 0} \\
K^{\star-} & K^{\star 0} & \phi \end{array}\right)^{\mu} \pm 
\frac{1}{ 2 \sqrt{2}}\left(\begin{array}{ccc}
\frac{f_{1 }+a_{1}^{0}}{\sqrt{2}} & a_{1}^{+} & K_{1}^{+} \\
{a_{1}^{-}} & \frac{f_{1 }-a_{1}^{0}}{\sqrt{2}} & K_{1}^{0} \\
K_{1}^{-} & \bar{K}_{1}^{0} & f_{1\phi}
\end{array}\right)^{\mu}
\end{eqnarray}

  Using the total effective Lagrangian of the model, we obtain the thermodynamic potential given as
\begin{equation}\label{totalt}
 \begin{array}{cc}
 \Omega\left(\sigma_{u}, \sigma_{d}, \sigma_{s},\omega, \rho, \phi, \Phi,\bar{\Phi}; T, \mu_{f}\right) = U\left(\sigma_{u}, \sigma_{d}, \sigma_{s}\right)+ 
 \Omega_{q \bar{q}}^{v a c}\left(\sigma_{u}, \sigma_{d}, \sigma_{s}\right)
 + \\
 \mathcal{U}\left(\Phi,\bar{\Phi}: T, \mu_{f}\right)+ 
 \Omega_{q \bar{q}}^{t h}\left(\sigma_{u}, \sigma_{d}, \sigma_{s}, \Phi,\bar{\Phi}: T, \mu_{f}\right) + V(\omega, \rho, \phi) -  U_0\left(\sigma_{u0}, \sigma_{d0}, \sigma_{s0}\right)
 \end{array}
 \end{equation}

 where the mesonic potential, including the chiral symmetry-breaking terms is described as \cite{stiele2014thermodynamics}
 \begin{eqnarray}\label{thermo}
 U\left(\sigma_{u}, \sigma_{d}, \sigma_{s}\right)= \frac{\lambda_{1}}{4}\left[\left(\frac{\sigma_{u}^{2}+\sigma_{d}^{2}}{2}\right)^{2}+\sigma_{s}^{4}+ 
 \left(\sigma_{u}^{2}+\sigma_{d}^{2}\right) \sigma_{s}^{2}\right]+ 
 \frac{\lambda_{2}}{4}\left(\frac{\sigma_{u}^{4}+\sigma_{d}^{4}}{4}+  
 \sigma_{s}^{4}\right)- \\  \nonumber
 \frac{c}{2 \sqrt{2}} \sigma_{u} \sigma_{d} \sigma_{s}+  
 \frac{m^{2}}{2}\left(\frac{\sigma_{u}^{2}+\sigma_{d}^{2}}{2}+\sigma_{s}^{2}\right)-
 \frac{h_{u d}}{2}\left(\sigma_{u}+\sigma_{d}\right)-h_{s} \sigma_{s} .
 \end{eqnarray}

  Since vector-like gauge symmetries are not spontaneously broken in a vacuum, $h_{ud}$ in the above equation represents the explicit symmetry breaking for the $u$ and $d$ quarks \cite{VAFA1984173}. The six parameters $\lambda_1$, $\lambda_2$, $m$, $c$, $h_{ud}$ and $h_s$  are determined by fitting the known decay constants $f_\pi$ and $f_K$ along with the masses of mesons such as $m_\pi$, $m_K$, $m_\sigma$ and squared masses of $\eta \prime$ and $\eta $ mesons \cite{lenaghan2000chiral}. At vanishing temperature and baryonic chemical potential values, the dynamical chiral symmetry breaking is taken into account by the fermion vacuum term in the model represented by $\Omega_{q \bar{q}}^{v a c}\left(\sigma_{u}, \sigma_{d}, \sigma_{s}\right)$ in Eq.~\ref{thermo}. Due to this term, the critical point is relocated to a lower temperature and higher baryonic chemical potential values \cite{gupta2012revisiting,chatterjee2012including}. Additionally, the vacuum potential energy, $U_0(\sigma_{u0}, \sigma_{d0}, \sigma_{s0})$ is subtracted to obtain the vanishing vacuum energy. The quark-antiquark interaction term is written as a combination of the vacuum mesonic fluctuations and thermal terms, $\Omega_{q \bar{q}}^{t h}$, which is deduced from the fermionic determinant. These terms can be described as \cite{PhysRevD.85.074018}
 \begin{equation}
 \Omega_{q\bar{q}}=\Omega_{q\bar{q}}^{vac}+\Omega_{q \bar{q}}^{t h} 
 \end{equation}
 where,
 \begin{equation}\label{equal6}
 \Omega_{q\bar{q}}^{vac}=-2 N_{c} \sum_{f=u,d,s} \int \frac{d^{3} p}{(2 \pi)^{3}} E_{f}^*
 =-\frac{N_{c}}{8 \pi^{2}} \sum_{f=u, d, s} m_{f}^{*^4} \log \left[\frac{m_{f}^*}{\Lambda}\right],
 \end{equation}
 
 \begin{eqnarray}\label{theq}
 \Omega_{q\bar{q}}^{th}=-2 T \sum_{f=u, d, s} \int \frac{d^{3} p}{(2 \pi)^{3}}\left[\ln {\bar g_{f}}+\ln g_{f}\right].
 \end{eqnarray}
 In above equation, $\bar g_{f}$ and $g_{f}$ are defined as 
 \begin{equation}
\bar g_{f}=\left[1+3 \Phi e^{-(E_f^*-\mu_f^*)/ T}+3 \bar{\Phi} e^{-2(E_f^*-\mu_f^*) / T}+e^{-3(E_f^*-\mu_f^*)/ T}\right],\\ 
 \end{equation}
 and
 \begin{equation}
 g_{f}=\left[1+3\bar\Phi e^{-(E_f^*+\mu_f^*)/ T} +3 \Phi e^{-2(E_f^*+\mu_f^*)/ T}  +e^{-3(E_f^*+\mu_f^*)/ T}\right].
 \end{equation}
 In Eq.~\ref{equal6}, $\Lambda$ is the regularisation scale parameter. The effective single particle energy of the quarks is modified due to the interactions with mesons and is defined as $E_f^{*}=\sqrt{p^2 + m_{f}^{*2}}$. The $m_f^*$ represents the effective mass of constituent quarks given by

 \begin{equation}
 m_u^{*} = \frac{g}{2}\sigma_u,\quad  m_d^{*} = \frac{g}{2}\sigma_d \quad \text{and} \quad m_s^{*} = \frac{g}{\sqrt{2}}\sigma_s.
 \end{equation}

The Yukawa coupling constant $g$ value is determined by fixing the mass of light quarks at $m_l$ = 300 MeV. In Eq.~\ref{totalt}, the interaction term incorporating the non-strange vector ($\omega$) and vector-isovector field ($\rho$), along with strange
vector field ($\phi$) is written as $V(\omega,\rho,\phi) = -\frac{1}{2}(m_{\omega}^{2}\omega^2 + m_{\rho}^2 \rho^2 + m_{\phi}^{2}\phi^2)$. The effective chemical potential of the quarks is modified as a consequence of vector-meson interactions \cite{KITAZAWA2003C289} and is defined in terms of quark chemical potential, $\mu_q$, isospin chemical potential, $\mu_I$ and strangeness chemical potential, $\mu_S$ as
 \begin{equation}\label{potential}
 \begin{array}{l}
 \mu_u^{*} =  \mu_q +  \mu_I - g_{\omega u} \omega - g_{\rho u}\rho    \\ 
 \mu_d^{*} =  \mu_q - \mu_I - g_{\omega d} \omega + g_{\rho d}\rho \\ 
 \mu_s^{*} =  \mu_q  -\mu_S  - g_{\phi s}\phi.
 \end{array}
 \end{equation}

 To study the properties of chiral symmetry breaking and deconfinement within the PQM framework, the model is extended by introducing gauge-invariant Polyakov loop potential. It is defined as the trace of a Wilson loop in the temporal direction, where the Wilson loop is a path-ordered exponential of the gauge field. In the infinite mass limit of quarks, the order parameter that contributes to confinement is determined by its expectation value. The Polaykov loop is defined in a manner that obeys the center symmetry \cite{polyakov1977thermal}. In the current work, we use the polynomial form of the Polyakov loop defined as \cite{grunfeld2018finite}

\begin{equation}
\frac{\mathcal{U}_{\text {poly }}(\Phi, \bar{\Phi})}{T^{4}}=-\frac{b_{2}(T)}{2} \bar{\Phi} \Phi-\frac{b_{3}}{6}\left(\Phi^{3}+\bar{\Phi}^{3}\right)+\\
\frac{b_{4}}{4}(\bar{\Phi} \Phi)^{2},
\end{equation}

and the temperature-dependent coefficient $b_{2}$ defined as
\begin{equation}
   b_{2}(T)=a_{0}+a_{1}\left(\frac{T_{0}}{T}\right)+a_{2}\left(\frac{T_{0}}{T}\right)^{2}+a_{3}\left(\frac{T_{0}}{T}\right)^{3} . 
\end{equation}
The parameters are determined by fitting the data to lattice simulations, which gives: $a_0=1.53$, $a_1=0.96$, $a_2=-2.3$, $a_3=-2.85$, $b_3=13.34$ and $b_4=14.88$ \cite{ratti2006phases}. To study the properties of different scalar and vector fields, the total thermodynamic potential in Eq.~\ref{totalt} is minimized with respect to the fields as
\begin{equation}
 \frac{\partial \Omega}{\partial \sigma_u} = \frac{\partial \Omega}{\partial \sigma_d} = \frac{\partial \Omega}{\partial \sigma_s} = \frac{\partial \Omega}{\partial \omega} = \frac{\partial \Omega}{\partial \rho} = \frac{\partial \Omega}{\partial \phi} = \frac{\partial \Omega}{\partial \Phi} = \frac{\partial \Omega}{\partial \bar {\Phi}}.
\end{equation}
Using the total thermodynamic potential, pressure density is calculated by the relation, $p$ = -$\Omega$. Further, the fluctuations of conserved charges are calculated by Taylor's series expansion method at zero value of the corresponding chemical potential. The susceptibilities of  $n_{th}$ order are written as
 \begin{equation}
    \chi^{qIS}_{ijk} =  \frac {\partial^{i+j+k} [p/T^{4}]} {\partial \left(\mu_{q}/T \right)^{i}
\partial(\mu_{I}/T)^{j} \partial(\mu_{S}/T)^{k}} 
\end{equation}

\begin{figure}
    \centering
    \includegraphics[scale=0.7]{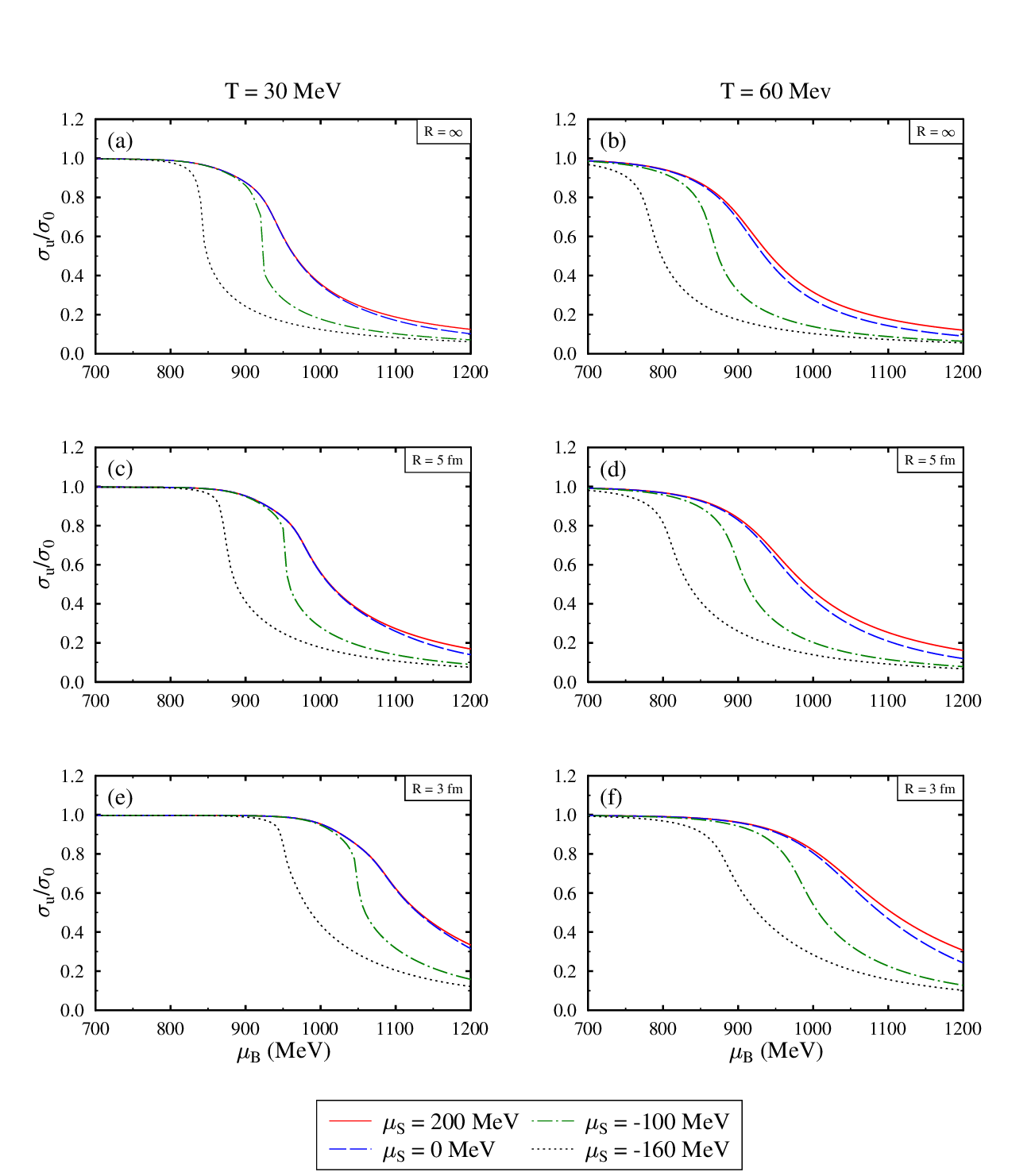}
    \caption{The up quark condensate plotted as a function of baryon chemical potential, $\mu_B$ for strangeness chemical potential, $\mu_S$ = 200, 0, -100 and -160 MeV, value of system size, $R$ = $\infty$, 5 and 3 fm and temperature, T = 30 and 60 MeV. }
    \label{figure1}
\end{figure}

\section{Results and discussion}  \label{results}

\begin{figure}
    \centering
    \includegraphics[scale=0.6]{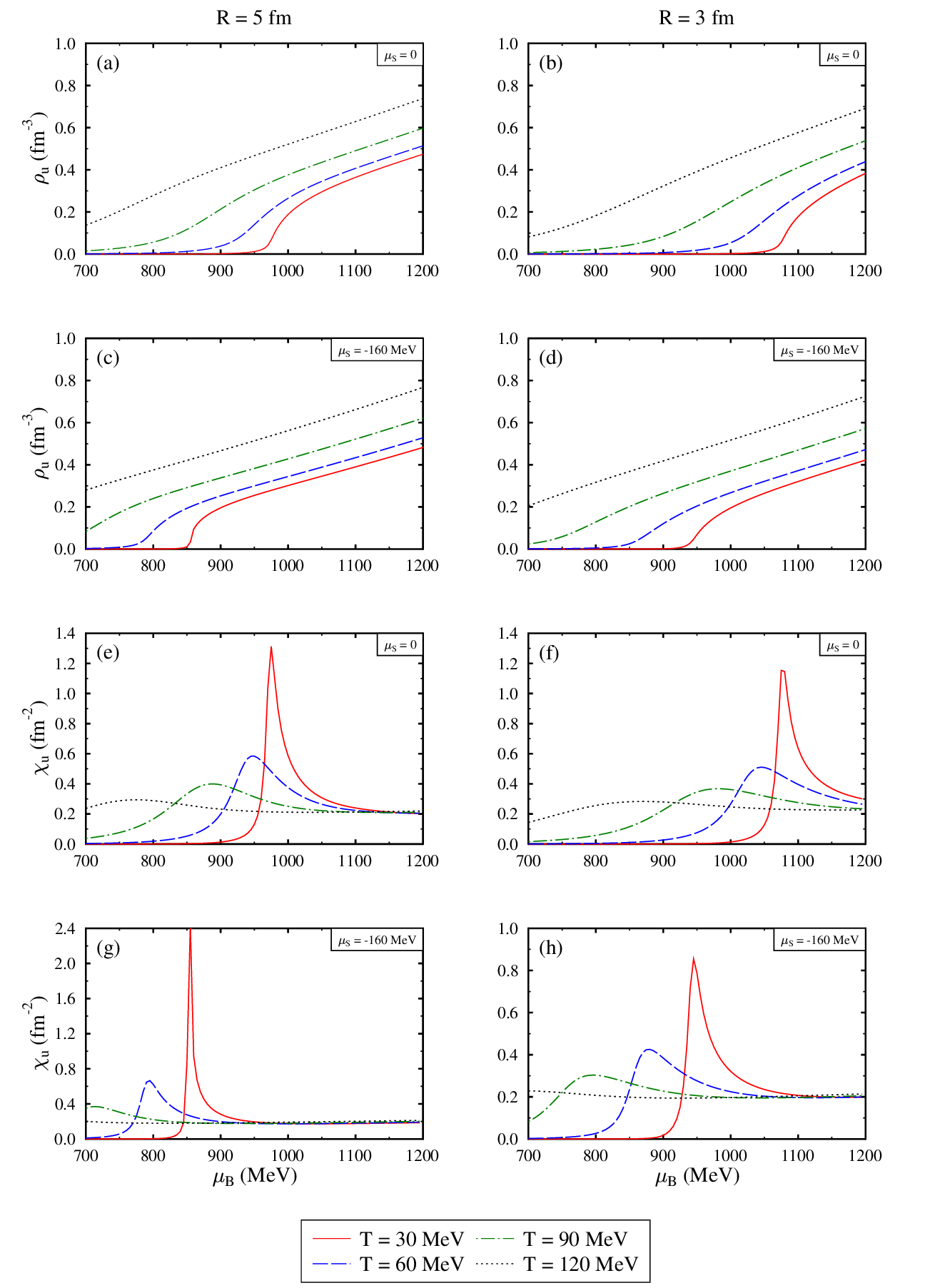}
    \caption{The up quark density, $\rho_u$ and susceptibility, $\chi_u$ plotted as a function of baryon chemical potential, $\mu_B$ for temperature, T =  30, 60, 90, 120 MeV, strangeness chemical potential, $\mu_S$ = 0, and -160 MeV and value of system size, $R$ = 5 and 3 fm. }
    \label{figure2}
\end{figure}

\begin{figure}
    \centering
    \includegraphics[scale=0.6]{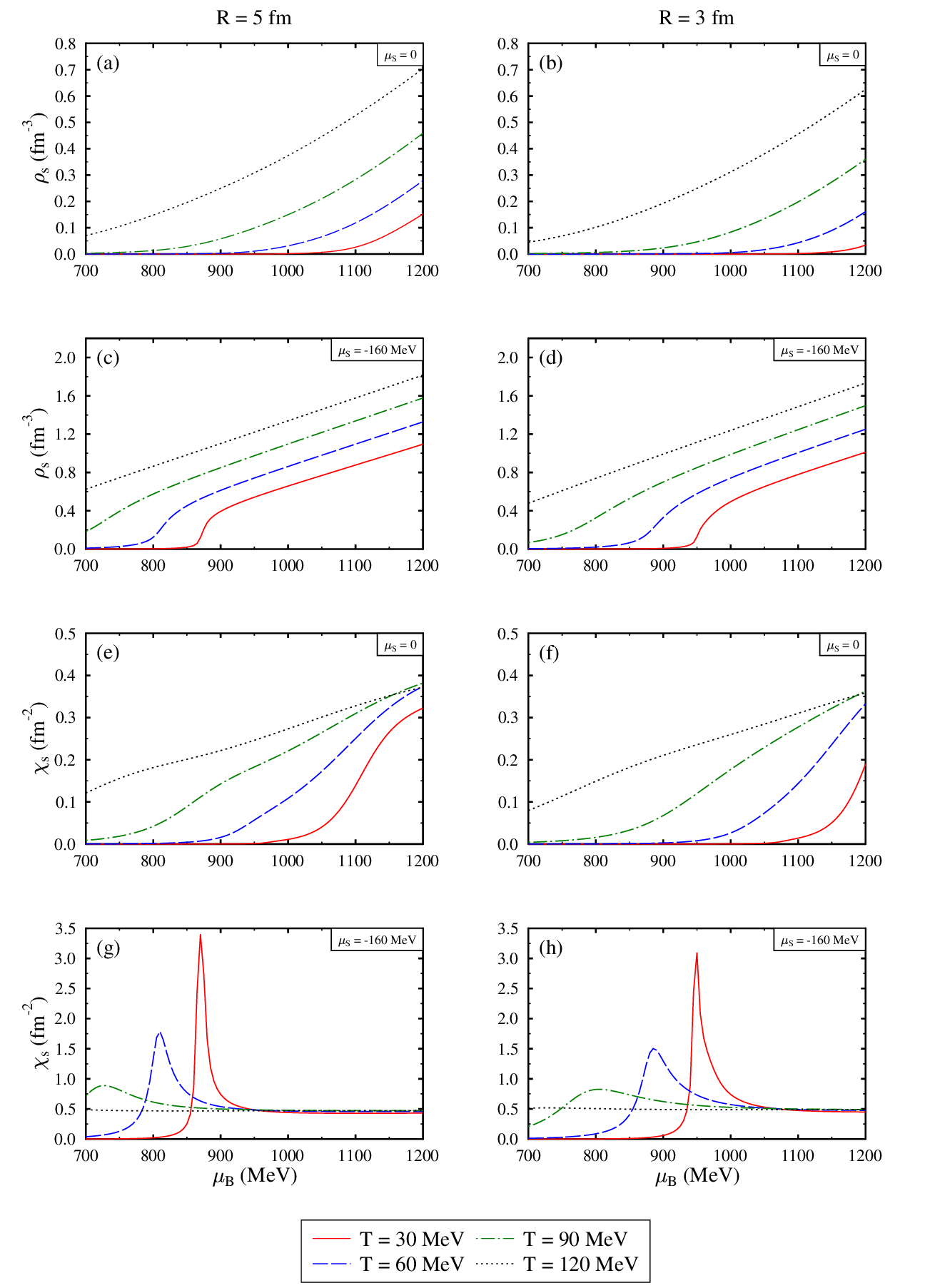}
    \caption{The strange quark density, $\rho_s$ and susceptibility, $\chi_s$ plotted as a function of baryon chemical potential, $\mu_B$ for temperature, T =  30, 60, 90, 120 MeV, strangeness chemical potential, $\mu_S$ = 0, and -160 MeV and value of system size, $R$ = 5 and 3 fm. }
    \label{figure3}
\end{figure}

\begin{figure}
    \centering
    \includegraphics[scale=0.7]{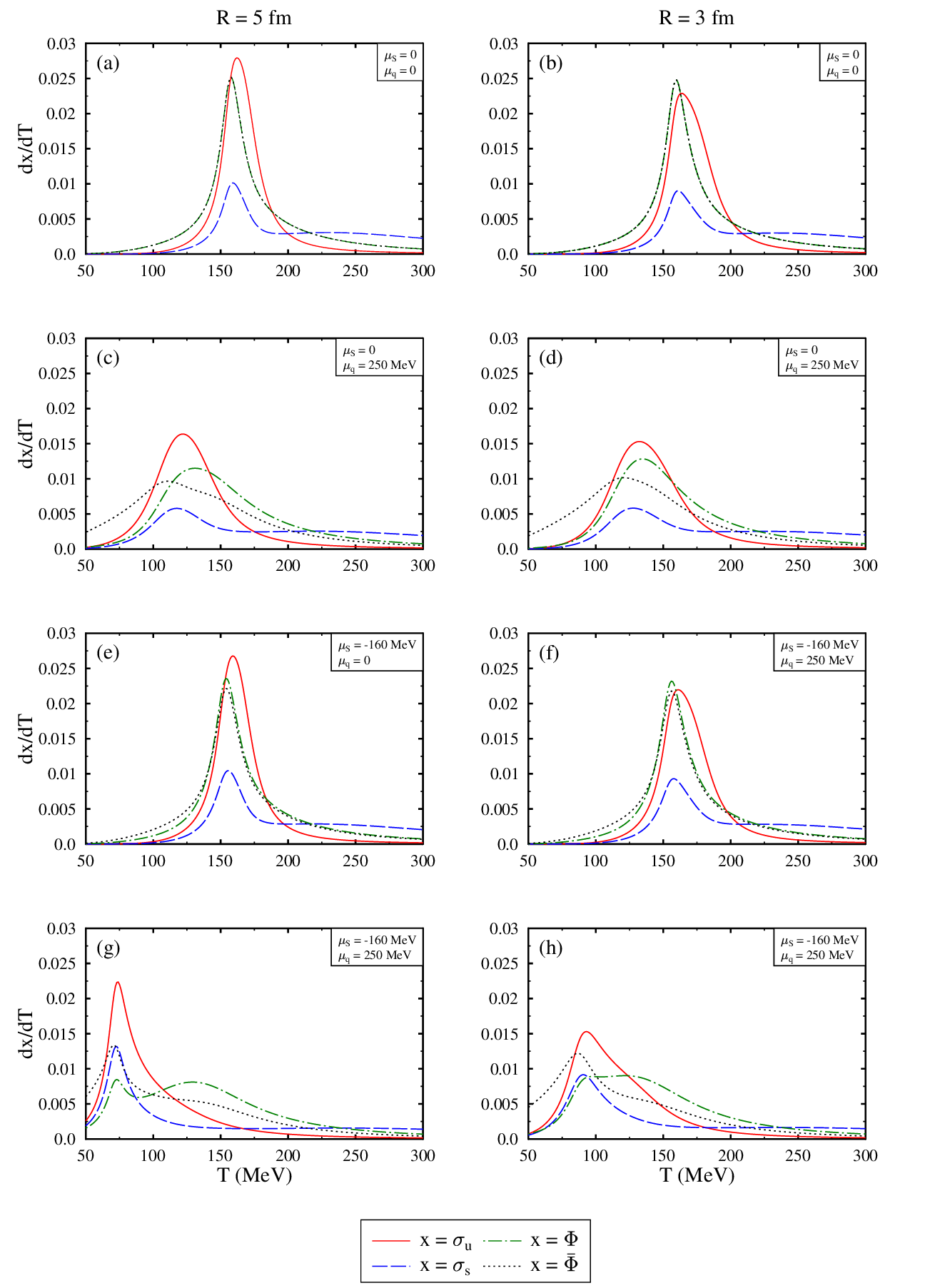}
    \caption{The derivatives of quark condensates and Polyakov loop variables plotted as a function of temperature, T for quark chemical potential, $\mu_q$ =  0 and 250 MeV, strangeness chemical potential, $\mu_S$ = 0, and -160 MeV and value of system size, $R$ = 5 and 3 fm.}
    \label{figure4}
\end{figure}

In this section, we have discussed the effect of the finite system size, $R$, and varying values of the strangeness chemical potential, $\mu_S$, on the asymmetric quark matter in the framework of the PQM model. The determination of the parameters of the model is dependent on the inclusion of vacuum mesonic fluctuations term. It has been highlighted that the expectation values of curvature mass and sigma field are used as input to calculate the sigma-meson mass, $m_\sigma$, and pion decay constant, $f_\pi$ \cite{2016}. In the current work, the value of the vector coupling constant has been fixed at $g_v$ = 6.5 and the isospin chemical potential, $\mu_I$ = 30 MeV for Figure~\ref{figure1} to Figure~\ref{figure4}. There is a finite value of isospin and strangeness chemical potential in heavy-ion collisions; thus the field values and thermodynamic quantities are calculated at finite values of these potentials \cite{braun1999chemical}.

In Figure~\ref{figure1}, we have shown the variation of up quark condensate, $\sigma_u/\sigma_0$, as a function of the baryonic chemical potential, $\mu_B$ for $R$ = $\infty$, 5 and 3 fm, and the strangeness chemical potential ranging from positive to negative for temperature values of T = 30 and 60 MeV. We have observed a sudden fall in the value of the quark condensate with increasing $\mu_B$, which signifies the phase change at higher chemical potential for all values of $\mu_S$ and $R$. For decreasing volume and at a given value of $\mu_S$, this change in chiral condensate values occurs at a higher value of $\mu_B$. 
 On the contrary, for a given system size and temperature, a drop in $\sigma_u/\sigma_0$ at lower baryonic chemical potential values for decreasing $\mu_S$ is observed. Hence, the critical chemical potential is found to be shifted towards higher values for decreasing system volume and increasing strangeness chemical potential. The nature of the phase transition from confined to deconfined state with increasing $\mu_B$ is more accurately predicted through derivatives of the strange and non-strange quark condensates.

The derivatives of the chiral condensates and the Polyakov loop parameters are calculated to study the QCD phase diagram and, hence plotting the chiral phase transition boundary and the deconfinement transition line, respectively. The first-order phase transition line ends at a critical point and becomes a crossover for the chiral limit \cite{stephanov2004qcd}. The deconfinement transition line remains a crossover for all the values of temperature. The first-order phase transition line is deduced by plotting the susceptibility of the chiral condensate as a function of baryonic chemical potential at a fixed temperature value. In Figure~\ref{figure2}, the variation in vector density, $\rho_u$ and susceptibility, $\chi_u$ of up quark has been shown as a function of the baryonic chemical potential for $R$ = 5 and 3 fm at $\mu_S$ = 200, 0, -100 and -160 MeV. The sharp rise in susceptibility values at low temperatures suggests the occurrence of first-order phase transition, while the smooth change signals the crossover. It is clear from Figure~\ref{figure2}(a) and (c) that the critical value of the baryonic chemical potential moves to a lower value at finite $\mu_S$ and Figure~\ref{figure2}(b) and (d) shows the shift to higher $\mu_B$ for decreasing volume. The sharp peak in the susceptibility of the up quark, at T = 30 MeV, confirms the first-order phase transition. The susceptibility peak smoothens with the increase in temperature values, indicating a crossover transition. The change in the phase transition line from crossover to first-order phase transition at varying values of $\mu_S$ and $R$ is highlighted by the peak of $\chi_u$. The curve becomes smoother, and very little change is observed at T = 120 MeV for all values of $R$ and $\mu_S$, showcasing the crossover regime.

  Figure~\ref{figure3} shows the dependence of the strange quark vector density, $\rho_s$, and susceptibility, $\chi_S$, on the variation of strangeness chemical potential and system size at different temperatures. For vanishing $\mu_S$ and given system size, $\rho_S$ is almost zero for $\mu_B$ $\approx$ 1050 MeV and T = 30 MeV. This shows that strange quarks are produced at the higher baryonic chemical potential for the vanishing value of strangeness chemical potential. With the increasing value of T and $\mu_B$, $\rho_S$ shows a monotonically increasing trend. For the finite value of $\mu_S$, the density of $s$ quarks in the system remains zero for finite $\mu_B$, and then a sudden change is observed at fixed volume. This sudden rise may signify the change in phase from a confined to a deconfined state. For high-temperature values, $\rho_S$ increases uniformly with $\mu_B$. The increase in density of $s$ quarks is observed to occur at a higher $\mu_B$ for $R$ = 3 fm at a given temperature and strangeness chemical potential value. In comparison to the up quark susceptibility in Figure~\ref{figure2}, no peak is observed for the vanishing value of $\mu_S$ at T = 30 MeV. Hence, the phase transition line of $s$ quark is a crossover at low-temperature values for given system sizes and zero $\mu_S$, which indicates that the phase transition line of strange quarks might not coincide with non-strange quarks at low-temperature values and vanishing strangeness chemical potential. The appearance of the peak in the $\chi_S$ at T = 30 and 60 MeV for both the given system volumes emphasizes the shift of critical temperature to a higher point in the QCD phase diagram.

 Figure~\ref{figure4} shows the derivatives of chiral strange and non-strange condensates and Polyakov loop variables with varying temperature values for a given quark chemical potential. The critical temperature is given by the peak of the derivatives plotted. For the vanishing value of $\mu_S$, there are only single peaks in the derivatives of chiral condensates and Polyakov loop parameters. For the zero value of $\mu_q$, the curve for $\Phi$ coincides with that of $\bar\Phi$ as these parameters have equal values in this confined regime. The shift of peaks for different values of $\mu_S$ and system size is the same as discussed earlier. For the finite value of $\mu_S$ and $\mu_q$, two peaks are observed for $\Phi$. In this scenario, the critical point is calculated by using the condition $\Phi$(T)/$\Phi(T \to \infty) >$ 1/2 \cite{mao2010phase}. 
 
 \begin{table}[b]
\centering
\begin{tabular}{|c|c|c|c|c|c|}
\hline
 $R$ value   &     $R$ = $\infty$  &   $R$ = 5 fm    &  $R$ = 3 fm & 
 $R$ = $\infty$  &
 $R$ = $\infty$  \\ \hline

 $\mu_S$ value   &      $\mu_S$ = 0   &    $\mu_S$ = 0     &    $\mu_S$ = 0  & 
 $\mu_S$ = -100 MeV   &
 $\mu_S$ = -160 MeV  \\  \hline
 
$T_{CP}$ (MeV)  & 55  & 46  &  37  & 83    & 87       \\ \hline
 $\mu_{q(CP)}$ (MeV)   &    306   &   321  &  352  & 268  &  238    \\ \hline

\end{tabular}
\caption{The value of critical temperature and quark chemical potential for finite values of system size and strangeness chemical potential.}
\label{table1}
\end{table}

\begin{figure}
    \centering
    \includegraphics[scale=0.7]{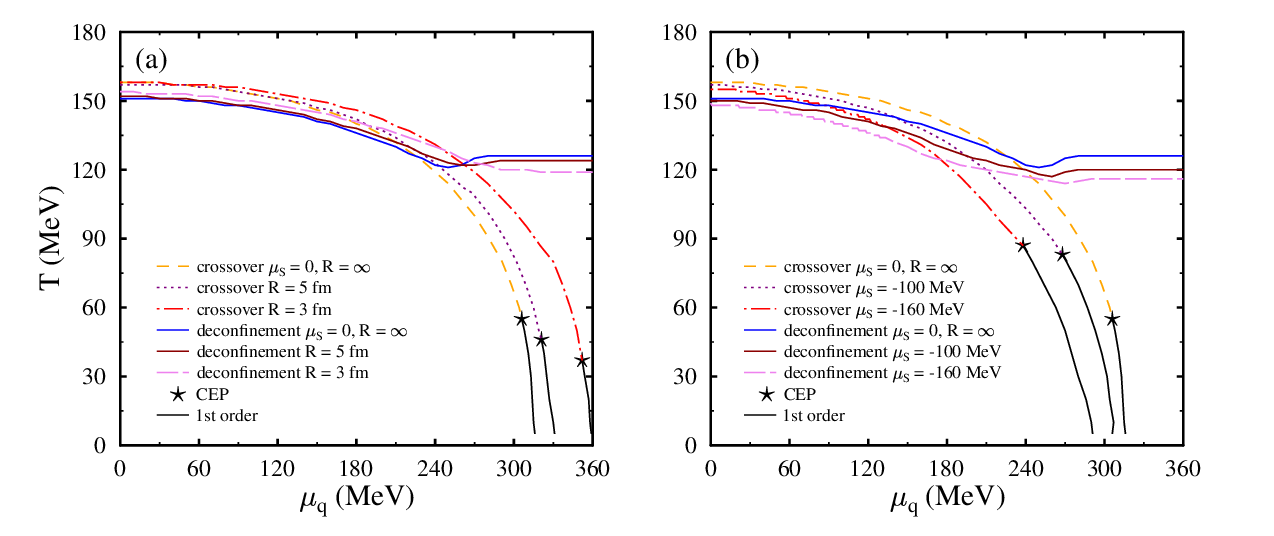}
    \caption{The QCD phase diagram for varying system sizes and strangeness chemical potential.}
    \label{figure5}
\end{figure}

 Using the critical temperature values derived from Figure~\ref{figure3} and Figure~\ref{figure4}, the QCD phase diagram for different system sizes and strangeness chemical potential has been displayed in Figure~\ref{figure5}. The chiral phase boundary and the deconfinement transition line have been shown for $R$ = $\infty$, 5 and 3 fm and $\mu_S$ = 0 in the right panel while for $\mu_S$ = 0, -100 and -160 MeV and $R$ = $\infty$ in the right panel. At the critical point, the first-order phase transition changes to the crossover for the chiral phase transition. The position of the critical point at the vanishing value of $\mu_S$ and infinite size is $(\mu^q_{CP}, T_{CP})$ = (306,55) MeV. For $R$ = 5 fm, the critical value of the temperature drops by 16.37 $\%$ and increases by 4.91 $\%$ for quark chemical potential. A further drop of 19.57 $\%$ for the critical temperature and a 9.66 $\%$ increase in quark chemical potential value is observed for $R$ = 3 fm. In the case of $\mu_S$ = -100 MeV, the critical temperature rises by 50.9 $\%$ whereas $\mu^q_{CP}$ falls by 12.42 $\%$. For $\mu_S$ = -160 MeV, critical quark chemical potential decreases by 11.2 $\%$, whereas $T_{CP}$ increases further by 4.81 $\%$. Hence, from the above discussion, it is clear that the critical point shifts to a lower temperature value and higher quark chemical potential with decreasing system size. A similar change in the value of the critical point has also been observed in the Polyakov loop modified Nambu-jona-lasinio (NJL) model \cite{tripolt2014effect,bhattacharyya2013thermodynamic}, though opposite change to a higher temperature for the critical point has also been reported with a decrease in volume in the framework of Polyakov chiral quark mean field model \cite{chahal2023effects}. The inclusion of the fermion vacuum fluctuation term in the current model significantly impacts the position of the QCD critical point \cite{chatterjee2012including,kovacs2023sensitivity}. As the value of $\mu_S$ approaches the positive value, the phase boundary becomes crossover with decreasing $T_{CP}$ and increasing $\mu^q_{CP}$. The change in the phase transition order at finite $\mu_S$ with increasing temperature has also been highlighted in \cite{toublan2005qcd,kumari2021quark}. The values of the critical points for varying volume and strangeness chemical potential has been listed in Table~\ref{table1}. The deconfinement boundary remains crossover for all temperatures and quark chemical potential values for all $R$ and $\mu_S$. A very small shift to lower values is observed for the deconfinement transition temperature for reduced system volume and negative values of strangeness chemical potential.

\begin{figure}
    \centering
    \includegraphics[scale=0.7]{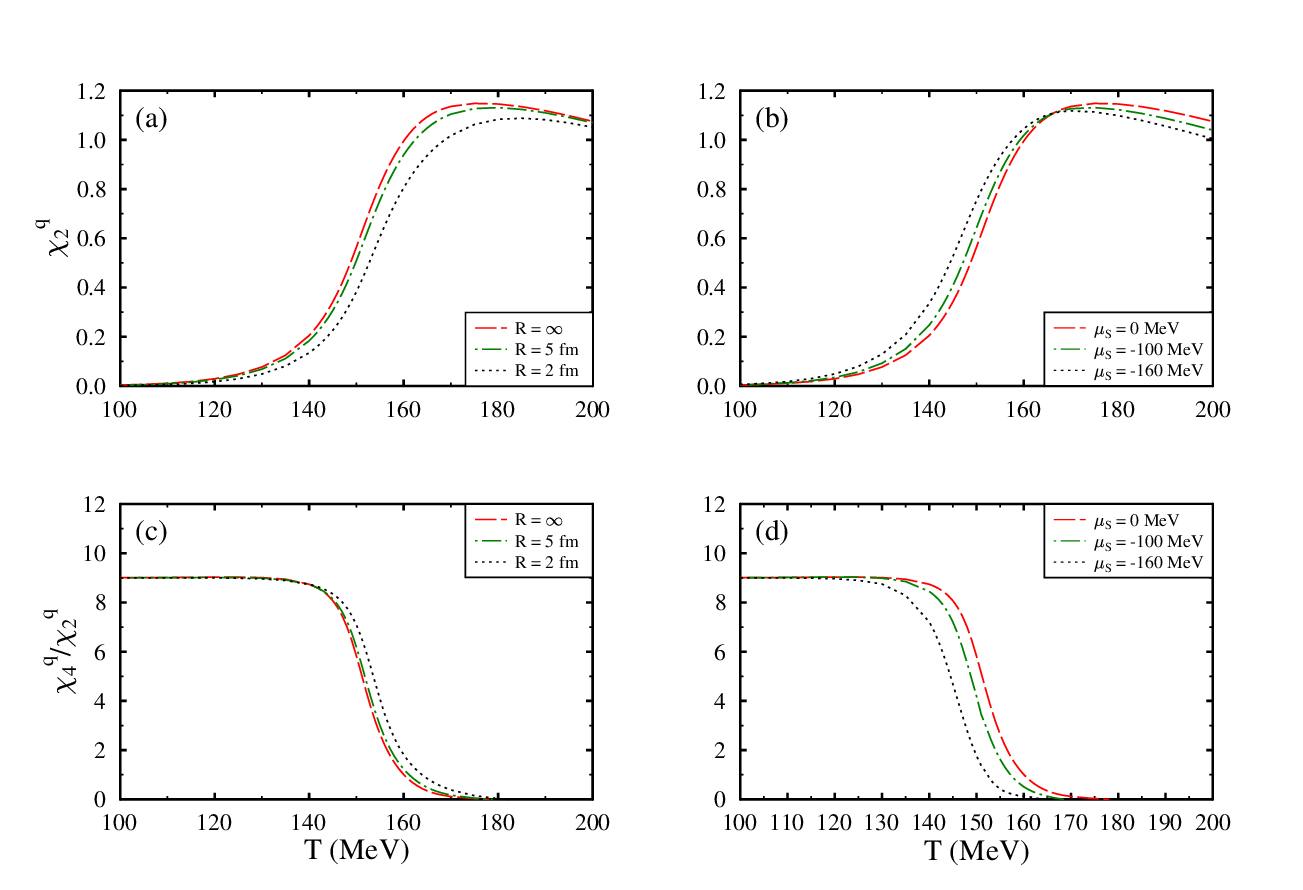}
    \caption{The second order susceptibility, $\chi_2^q$ and kurtosis, $\chi_4^q$/$\chi_2^q$  plotted as a function of temperature, T for strangeness chemical potential, $\mu_S$ = 0, -100 and -160 MeV and value of system size, $R$ = $\infty$, 5 and 3 fm.}
    \label{figure6}
\end{figure}

\begin{figure}
    \centering
    \includegraphics[scale=0.7]{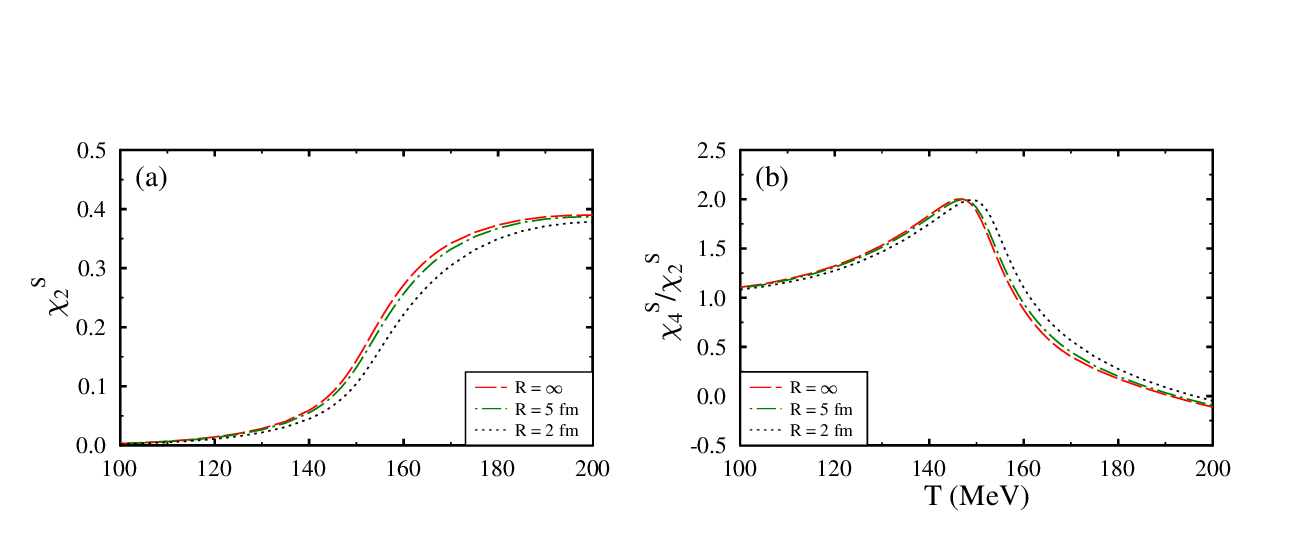}
    \caption{The second order susceptibility, $\chi_2^S$, and kurtosis, $\chi_4^S$/$\chi_2^S$  plotted as a function of temperature, T for the value of system size, $R$ = $\infty$, 5 and 3 fm.}
    \label{figure7}
\end{figure}

Fluctuations and susceptibilities of conserved charges have been recognized as observables, which helps to find the location and nature of the QCD critical point. Figure~\ref{figure6} and Figure~\ref{figure7} show the second-order susceptibility and kurtosis of quark number and strangeness number for varying values of system volume and $\mu_S$. The susceptibilities have been calculated using Taylor's series expansion method for vanishing values of corresponding chemical potentials. In Figure~\ref{figure6}(a) and (c), for zero value of $\mu_S$ and changing system size, $\chi_2^q$ and $\chi_4^q/\chi_2^q$ changes rapidly near the transition regime. There is a slight shift in the critical temperature value to a higher temperature with decreasing system size. But as discussed earlier, for finite values of $\mu_q$, the critical temperature shifts towards lower values for a reduced system volume.
 On the other hand, for finite values of $\mu_S$ and infinite system size, the critical temperature is shifted to a lower value for decreasing $\mu_S$, which contrasts with results discussed at finite $\mu_q$. The kurtosis value drops nearly to zero for higher temperatures, which signifies the change in degrees of freedom to quarks. The value of second-order susceptibility for quarks increases monotonically with the rise in temperature in both scenarios. The critical temperature for zero value of $\mu_q$ appears to be around 155 MeV, which coincides with the results of lattice QCD \cite{DING201452}. The value of susceptibilities depends on the form of the Polyakov loop under consideration and vector interactions. The comparison of these fluctuations of conserved charges with lattice data for changing $g_v$ and the Polyakov loop has been discussed in earlier work for the zero value of $\mu_S$ and infinite volume \cite{chahal2022quark}. 

The trend of second-order susceptibility of strangeness number is similar to that of $\chi_2^q$. The kurtosis for strangeness number shows a peak around the transition regime. The peak position is found to be shifted towards a lower temperature value for increasing system volume. This is similar to the change observed in susceptibilities of quark number. While investigating the derivatives of the chiral condensates, the change observed in critical temperature is opposite to that studied for finite chemical potential value.

\section{Summary} \label{summary}
To summarize, we have discussed the thermodynamic properties of asymmetric quark matter using the Polyakov quark meson model extended by introducing vector interactions. The effects of strangeness chemical potential, and finite system size have been investigated by analyzing the variation of strange and non-strange fields at varying temperatures and densities. The derivatives of quark condensates and Polyakov loop variables have been studied to locate the position of the QCD critical point. The impact of strangeness chemical potential has been highlighted by studying the strangeness fraction for zero and the negative value of $\mu_S$ at different temperatures. We have observed that the production of s quarks gets saturated at high-temperature values. The chiral phase boundary for $u$, $d$, and $s$ quarks coincides with a crossover transition at lower $\mu_q$ before the critical point and a first-order phase transition at higher values of quark chemical potential. The deconfinement phase boundary remains a crossover for all temperature values. With decreasing system volume, the critical point is found to shift to lower values of temperature and higher values of quark chemical potential. The positioning of the critical point is affected by the inclusion of the vacuum term in the model. On the other hand, the critical point is repositioned to a higher temperature and lower values of quark chemical potential for the decreasing value of strangeness chemical potential. The susceptibilities of conserved charges are enhanced in the transition region. The peak of the kurtosis for the strangeness number gives the value of the critical point to be $\approx$ 155 MeV, which is consistent with the lattice QCD studies at zero chemical potential. In future work, the susceptibilities of conserved charges can be studied at the finite value of chemical potential \cite{fan2019second}. The model can be further improvised using the functional renormalization approach \cite{dupuis2021nonperturbative, fu2022qcd}.

 \section*{Acknowledgment}

The authors sincerely acknowledge the support for this work from the Ministry of Science and Human Resources (MHRD), Government of India, through an Institute fellowship under the National Institute of Technology Jalandhar. Arvind Kumar sincerely acknowledges the DST-SERB, Government of India, for funding research project CRG/2019/000096.

\bibliographystyle{elsarticle-num} 
\bibliography{bib}

\end{document}